# Neurophenomenal Structuralism and the Role of Computational Context

**Marlo Paßler**                                        marlo.passler@hu-berlin.com

**Affiliations**:

**Berlin School of Mind and Brain**, Humboldt-Universität zu Berlin, Germany

**RTG 2386 "Extrospection"**, Humboldt-Universität zu Berlin, Germany

**Otto-von-Guericke University**, Institute of Philosophy, Magdeburg, Germany

Funded by the Deutsche Forschungsgesellschaft (DFG, German Research Foundation) - RTG 2386

**Adrien Doerig**

**Affiliations**:

**Freie Universität Berlin**, Cognitive Computational Neuroscience lab, Germany

Funded by ERC starting grant 101039524 TIME

**Abstract:** Neurophenomenal structuralism posits that conscious experiences are defined relationally and that their phenomenal structures are mirrored by neural structures. While this approach offers a promising framework for identifying neural correlates of contents of consciousness (NCCCs), we argue that merely establishing structural correspondences between neural and phenomenal structures is insufficient. This paper emphasizes the critical role of computational context in determining the content of neural structures. We introduce four criteria—Sensitivity, Organization, Exploitation, and Contextualization—to evaluate which neural structures are viable NCCC candidates. These criteria highlight that, for neural structures to meaningfully mirror phenomenal structures they have to be actively exploited and be able to influence behavior in a structure-preserving way. Our analysis demonstrates that anatomical and causal neural structures fail to meet certain criteria, whereas activation structures can, provided they are embedded within the appropriate computational context. Our findings challenge both local and rich global structuralist theories for overlooking the content-constituting role of computational context, leading to proposed NCCCs that fail to fully account for conscious content. We conclude that incorporating computational context is essential for any structuralist account of consciousness, as it determines the nature of dimensions within neural activation spaces and, consequently, the content of conscious experiences.

# 1. Introduction

In the ever-evolving field of consciousness science, an approach known as 'neurophenomenal structuralism' is making waves. Championed by scholars such as Fink et al. (2021), Tsuchiya and Saigo (2021), Lau et al. (2022), Lyre (2022), and Kob (2023), this approach offers a fresh perspective on the relationship between neural states and conscious experiences. Central to this approach is the assertion that one cannot characterize a conscious experience in isolation. For example, we cannot characterize the experience of blue. However, we can characterize the relationships between experiences. Roughly, blue is perceptually more similar to purple than to orange. Further, neurophenomenal structuralists claim that this phenomenal structure is mirrored in neural structure. Simply put, they propose that the neural vehicles of blue percepts are more similar to the neural vehicles of purple percepts than to those of orange percepts. As such, experiences are relationally defined, and the resulting similarity structures are systematically mirrored by their neural vehicles. This neural-to-phenomenal (henceforth 'neurophenomenal') structural mapping is offered as a means to pinpoint the neural substrate of consciousness, or at least its content-constituting parts.

This approach diverges significantly from traditional methods that aim to isolate the neural correlate of consciousness (NCC) by contrasting conscious and unconscious conditions in paradigms where stimuli are manipulated (e.g. using masking) to ensure that they are processed unconsciously in certain trials (Baars 1986). Neurophenomenal structuralism, instead, does not rely on the difficult distinction between conscious vs. unconscious processing (Irvine 2012; Overgaard 2015; Phillips 2018; Michel 2020; Irvine 2021; Paßler 2023). Rather, it proposes to consider similarity relations between conscious stimuli of a given experiential domain (e.g. colors), and to search for neural underpinnings that share that structure. Here, we aim to identify criteria that make a neural structure a plausible candidate for mirroring phenomenal structure, as suggested by neurophenomenal structuralism.

First, we stress that subjective reports play an unavoidable and central role in measuring phenomenal structure experimentally. Any structuralist theory that aims to be supported empirically needs to contend with this fact. Having established that the phenomenal side of neurophenomenal mappings must be measured using behavioral proxies, we turn to the neural side.

Our central argument is that, in order to be reliably mapped onto phenomenal structures, neural structures must systematically influence the behaviors used in psychophysics to approximate these phenomenal structures. We consider different types of neural structures—anatomical, causal, and activation structures—and discuss which of these can impact behavior in a structure-preserving manner. We argue that neither anatomical nor causal structures can systematically impact behavior, making them unlikely candidates for a neurophenomenal mapping. On the other hand, activation structures can systematically impact behavior, given certain conditions.

The criteria we establish along the way underscore the critical role of computational context. More specifically, for an activation structure to contribute meaningfully to subjective reports, the relations it embodies must hold computational relevance for downstream brain processes. Further, this computational context is essential for determining the content of these structures. Without it, activation structures remain highly ambiguous and fail to specify the content of an experience. We conclude that, within a structuralist framework, local accounts are deemed to fail because they overlook the crucial contribution of computational context in shaping the content of consciousness.

By including the computational context in the scope of inquiry, we aim not just to critique current approaches, but rather to deepen and enrich the ongoing debate in neurophenomenal structuralism.

## 2. Phenomenal Structure and the Role of Subjective Reports

According to neurophenomenal structuralism, our experiences are best understood through their relationships with other experiences. For instance, the experience of seeing red cannot be described in isolation, but only in relation to similar colors like orange and pink, as well as in contrast to dissimilar colors like blue or green. These comparative statements provide a practical and relatable way to convey what an experience feels like. Based on these comparative statements, we can construct abstract 'quality spaces' (Clark 1993; Rosenthal 2010, 2015), that describe these (dis)similarities between experiences.

In this section, we provide a brief history of how psychophysicists have attempted to construct quality spaces to quantify the structure of experiences. In doing so, we emphasize the importance of subjective reports for structuralist approaches to consciousness. A prime example of this approach is the color space (see figure 1).

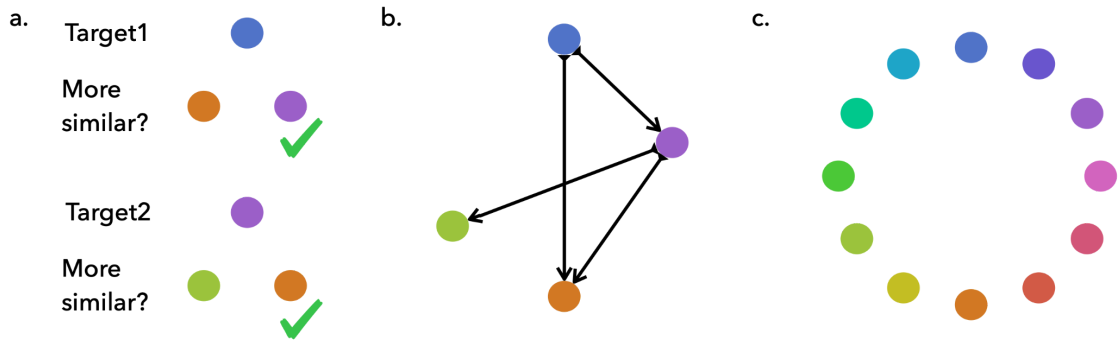

**Figure 1**: From Subjective Reports to Quality Spaces
a.) Participants provide subjective color similarity reports, for example by reporting which of two colors is more similar to a target color. b.) These similarity reports are translated into spatial distances, with colors judged to me more similar placed closer together. c.) Conducting these steps over a large number of stimuli allows researchers to construct quality spaces. In the case of colors, the quality space is ring-like.

The construction of the color space starts with psychophysical experiments. Participants engage in tasks where they evaluate and report on the subjective similarity between various colors, non-verbal reports included. Such tasks might prompt participants to consider which color, e.g. purple or orange, more closely resembles a reference color, e.g. blue. Their responses—indicating, for example, that purple is subjectively closer to blue than orange—are then quantitatively translated into spatial proximities (using algorithms such as Mulit-Dimensional Scaling). In the simple example of Fig. 1, colors judged to be more alike are placed in closer proximity, while those deemed dissimilar are spaced further apart. When conducted carefully to cover the range of visible colors, the endpoint of this process is an extensive quality space for color that encompasses the entire spectrum of colors with the relative distances between them corresponding to their perceived similarities (Churchland 2005).

Quality space theory and psychophysics more generally have revealed robust and intriguing peculiarities of the color space, such as its circular nature, noted already by Newton in 1704 and probably even Aristotle (Fig. 1C; see Parkhurst and Feller (1982)). Note that the physical wavelength spectrum is linear, in contrast to the circular human color space, which shows that quality spaces may not mirror a corresponding physical space (Ramanath et al. 2004). Another interesting feature of the human color space is its asymmetry in the perceivable brightness of hues, where blues can appear darker than yellows, and vice versa (Palmer 1999). As a result, the color space emerges as a circular and asymmetric cone-shaped structure, showcasing the distinctive insights quality space theory can provide.

Crucially, this space and all its surprising characteristics is derived from careful experiments involving subjective similarity ratings about colors (Fig. 1a) and not purely from first person introspection.

Such explorations of quality spaces derived from subjective reports have been fruitful across various sensory domains—from pitch and taste to odor and others. Each of these analyses uncovered distinct structures. Pitch space, for instance, was visualized as a helix (Shepard 1982), taste space as a tetrahedron (Henning 1916b; Gärdenfors 2004), and early models linked odor to a prism structure (Henning 1916a); Fig. 2). Despite these achievements, inferring quality spaces from subjective reports can be challenging.

Henning's (1916a) exploration into constructing an odor space is a good example of the complexities that can arise. Henning's approach involved identifying a number of primary odors and understanding their interrelationships. He based his investigation on the assumption of six pure primary odors, based on qualitative descriptions by Henning himself and two collaborators during a walk through a botanical garden (Wise et al. 2000). Among other things, participants were presented with sequences of unlabeled odors and tasked with aligning them linearly according to their similarities, while identifying and removing outliers. This approach was intended to reveal the natural progression from one primary odor (the odors regularly found at the end of the ordered lines) to another, without explicitly naming the odors. Through this process, Henning observed that transitions between primaries did not involve intermediary primaries, but were smooth progressions of primary mixtures, either direct transitions of just two primaries (Fig.2; a transition along an edge, e.g., spicy-to-flowery) or complex transitions involving more than just two primaries (Fig.2; a transition through the body of the figure, e.g., spicy-to-fruity). Such observations led him to favor the prism model.

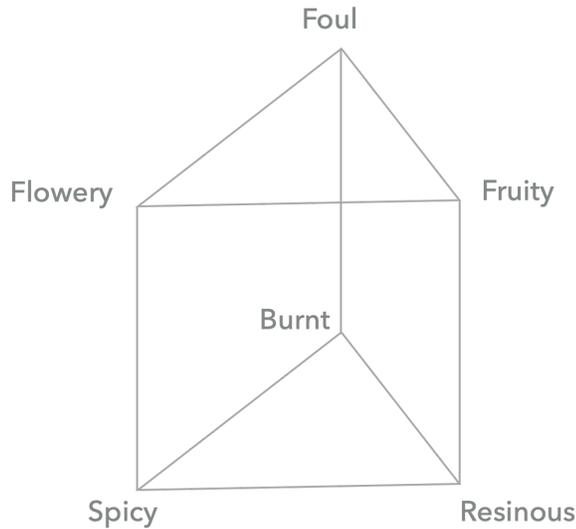

**Figure 2**: Henning's (1916) odor prism translated from German.

However, the methodology behind Henning's construction of the odor prism faced criticism, in particular for its reliance on qualitative assessments from a fairly limited number of subjects. Dimmick (1927) pointed out the absence of robust statistical validation in Henning's identification of the six primary odors. The presumption of six primaries naturally biases the analyses of the qualitative data on transitions between them. Hence, while Henning's odor prism model presented an intuitively appealing representation of olfactory space, acknowledging direct and complex odor transitions and more, it lacked statistical rigor. The number of primaries and dimensions of odor space still remains controversial (Bushdid et al. 2014; Meister 2015).

To overcome such caveats, researchers have continuously developed more sophisticated methods to analyze subjective reports, and to address the many biases and variabilities inherent in these reports (Dimmick 1927; Lawless 1989; MacRae, Howgate, and Geelhoed 1990). Techniques such as direct pairwise similarity ratings (Yoshida 1964) and Multidimensional Scaling (MDS) (Mead 1992) have significantly advanced the field by enabling more quantitative analyses of similarity. Recent innovations, such as "inverse MDS" (Kriegeskorte and Mur 2012) and feature-weighted representational similarity analysis (Kaniuth and Hebart 2022), further refined our understanding by factoring in inter-individual differences and contextual dependencies, which were previously considered highly problematic (Tversky 1977). The integration of machine learning has further strengthened data-driven approaches and greatly increased the volume of subjective ratings we can handle (Lötsch et al. 2019).

This comprehensive evolution from few qualitative assessments to increasingly refined data-driven methods reflects a profound and ever-improving approach toward understanding and reconstructing the organization of odor space through behavioral data. These developments enhance the statistical power and robustness of our quality spaces suggesting that we are progressively capturing meaningful aspects of the underlying phenomenal structure. The only constant in this evolution is the persistent need for behavioral proxies and subjective reports as critical benchmarks to approximate phenomenal structure. This underscores the report-based nature of quality space theory and, thus, of the phenomenal side of the neurophenomenal mapping.

The primacy of subjective reports as a source of information about consciousness is generally agreed-upon. Recent discussions have emphasized this as a significant constraint on our methods of verification regarding theories of consciousness (Doerig et al. 2019; Doerig, Schurger, and Herzog 2021; Herzog, Schurger, and Doerig 2022). Herzog et al. (2022) distinguish the pure first-person experience from its subjective third-person reports, noting that pure first-person experiences, being inherently private, cannot support or refute scientific theories as they can only be referenced through their behavioral third-person surrogates. This underscores the critical role of subjective reports in mapping out phenomenal structure, a standpoint even acknowledged by staunch advocates of 'first-person' methodologies like Tsuchiya et al. (2019): "In consciousness science we rely on the link—known to each of us by first-hand experience—between conscious experience and behavioral reports. This link motivates our collection of physical measurements including behavioral reports, and motivates theories about consciousness."

From this perspective, it is crucial to articulate in precise terms that what we investigate is not directly a neuro-phenomenal mapping but a neuro-behavioral mapping. Quality spaces, as we can conceptualize them, are abstract spaces derived from behavioral data that stand in for phenomenal structure. This fact doesn't diminish the utility or significance of quality spaces in exploring the dimensions of human experience; rather, it underscores a critical methodological nuance. Quality spaces are measurable proxies for phenomenal structure—a behavior-based approximation that, while not capturing subjective experience directly, provides a practical, accessible, and possibly the only means of investigation.

## 3. Beyond Correspondence: The Importance of Exploitability in Identifying NCCCs

The previous section highlighted the role of subjective reports and their derivatives—quality spaces—as indispensable proxies for the otherwise elusive phenomenal structure of consciousness. We argue next that a key assumption for neurophenomenal structuralism must be that the neural structures mirroring phenomenal structures also shape subjective reports in a structure-preserving fashion. In the following sections, we show how this insight can help specify criteria that a neural structure must meet in order to be promising for the neurophenomenal structuralist agenda. The argument has three premises and a conclusion:

*Premise 1:* There is a structural mapping between neural and phenomenal structures (neurophenomenal structuralism assumption).

*Premise 2:* Under standard conditions, there is a systematic relation between the structure provided by similarity ratings and phenomenal structure (measurability assumption).

*Premise 3:* The systematic relationship described in Premise 2 can only be accounted for by Premise 1 if the neural structures that mirror phenomenal structure also influence behavior in a structure-preserving way.

*Conclusion:* Therefore, for neurophenomenal structuralism to be empirically relevant, the neural structures that mirror phenomenal structures must impact behavior in a structure-preserving way.

Premise 1 is the foundational assumption of neurophenomenal structuralism. Questions arise on both sides of this mapping assumption. Some researchers debate whether the neural vehicles of perception are genuinely structural or if they rely on a more discrete format (VanRullen and Koch 2003; Herzog et al. 2020; Tee and Taylor 2020; Quilty-Dunn et al. 2022). Others question whether conscious experience is as richly structured as it appears (Kouider et al. 2007; Kouider et al. 2010; Cohen et al. 2016; Kammerer 2021). We set aside these and other potential objections (Fink and Kob 2023; Kleiner 2024) to focus on the implications of accepting it.

Premise 2 builds on the earlier methodological nuance: we can only measure phenomenal structure through behavioral proxies. Since consciousness cannot be measured directly from a first-person perspective, our best approach is to gather public data grounded in subjective reports—i.e., behavior—under standard conditions (see Feest 2014; Overgaard 2015; Michel and Lau 2020; Pauen and Haynes 2021; Herzog et al. 2022; Paßler 2023). Therefore, we must assume that the quality spaces we derive from behavior are good measurable proxies of phenomenal structure.

Of course, this assumption doesn't hold under all conditions. For instance, if a participant is coerced or biased, the structure of their subjective reports might not reflect their actual experience. Similarly, locked-in patients may be unable to provide reports, though this doesn't imply their phenomenal structure is void. However, under normal conditions, it is both natural and necessary to assume premise 2 (Pauen and Haynes 2021; Paßler 2023).

Premise 3 bridges the first two premises. It asserts that the systematic relationship between phenomenal structure and subjective reports (Premise 2) is valid only if the neural structures that mirror phenomenal structure (Premise 1) influence behavior in a structure-preserving way. More specifically, for subjective reports to accurately reflect phenomenal structure under neurophenomenal structuralism, the mirroring neural structures must shape them in a way that preserves structural integrity. Otherwise, the link between neural and phenomenal structures would not reliably extend to the behaviors we use in approximating phenomenal structure.

We conclude that an empirically testable neurophenomenal structuralism must assume that neural structures reflecting phenomenal structures influence subjective reports in a structure-preserving way. This is the only way to ensure that the quality spaces derived from these reports accurately depict the neural structures that mirror phenomenal structure. If this condition is not met, quality spaces can no longer serve as reliable proxies for either the neural or phenomenal structures.

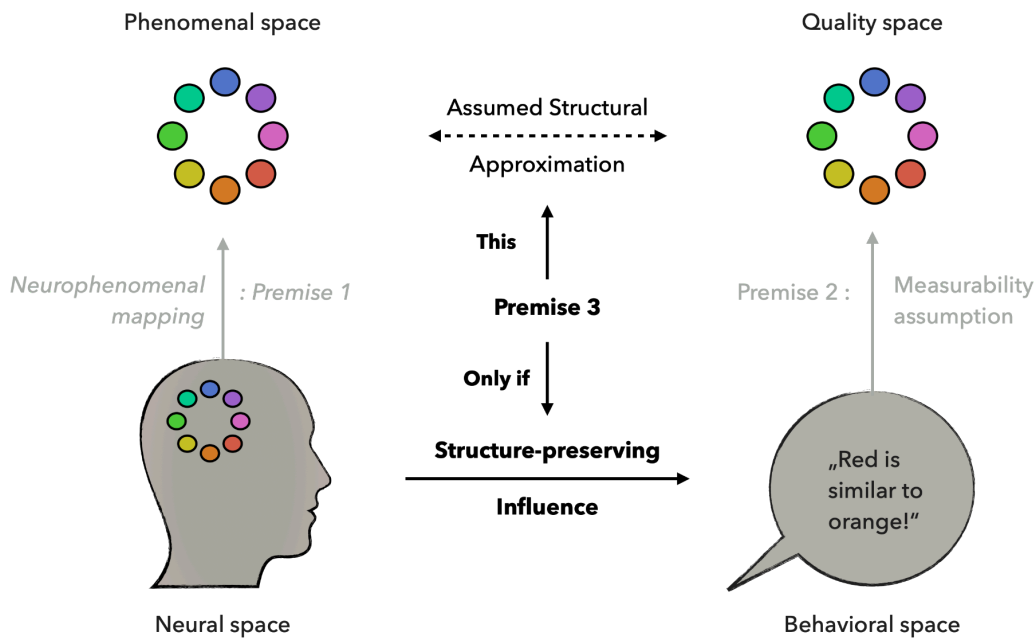

**Figure 3**: Illustration of the indirect route
The left side shows Premise 1, the mapping assumption between neural and phenomenal structures. The right side shows Premise 2, where behaviors reflect similarities across experiences (e.g., 'Red is similar to orange'). In the center, Premise 3 highlights that this structural approximation is valid only if neural structures mirroring phenomenal structures also shape behavior in a structure-preserving way.

In summary, our exploration of neurophenomenal mappings follows an indirect approach (Fig. 3). Instead of aiming to directly understand how neural structures give rise to phenomenal structures, we propose investigating how neural structures influence the behaviors that underpin the psychophysical analyses of quality spaces.

But how can we determine whether a neural structure reliably impacts behavior in a structure-preserving fashion? This is a challenging question, particularly because the brain is a complex and interconnected system with spurious structural resemblances. Putnam's (1981) famous example illustrates this issue: an ant's trail might accidentally resemble Winston Churchill, but this resemblance is purely coincidental and devoid of any significance. There is no causal interaction between the ant's path and Churchill's appearance; the ant's trail did not form as a result of Churchill's features, nor would a change in the ant's path have any effect on Churchill's appearance.

This scenario is linked to a key question in the structural representation literature: how to differentiate trivial structural correspondences from genuine structural representations. This problem can be addressed by distinguishing trivial correspondences (such as the ant's path) from cases where the corresponding structure can be actively *exploited* by a system (Ramsey 2007; Shea 2014, 2018; Gładziejewski and Miłkowski 2017). It is generally assumed that for a neural structure to be considered exploitable, its relational configuration must be systematically usable by the system's downstream processes.

This concept of exploitability from the literature on structural representations is closely related to our third premise. Exploitability ensures that a given neural structure can influence behavior in a structure-preserving way, and thus that quality spaces derived from these behaviors can be used as proxies for phenomenal structures (given the neural structures mirror them).

Our approach differs from standard neurophenomenal structuralist approaches in that it gives a central role to the *exploitability* of neural structures, instead of only focussing on their degree of *correspondence* with phenomenal structures. The idea behind the latter approach is to first identify a phenomenal structure and then "scan the brain" to find the neural structure that best matches this phenomenal structure (Fink et al. 2021; Tsuchiya and Saigo 2021). There, the degree of correspondence is a primary criterion for identifying Neural Correlates of Contents of Consciousness (NCCCs). However, according to our reasoning, the best-fitting neural structure may not be exploitable in the right way, necessitating its exclusion as a *structural* NCCC candidate (henceforth just "NCCC candidate"). This is where the concept of exploitability becomes crucial.

In the next section, we clarify this point by demonstrating how neural structures that fail to meet our exploitability-related criteria can easily be dissociated from behaviorally measured quality spaces. Such dissociations undermine the idea that these quality spaces reliably reflect the respective neural structures and, thus, the phenomenal structures they potentially mirror. This logic will be applied throughout the rest of this paper to help neurophenomenal structuralists ensure they are capturing something truly meaningful and empirically grounded.

## 4. Which kinds of neural structures are promising NCCC candidates?

"Neural structures" are central to the neurophenomenal structuralism approach, but there are various types of neural structures within the brain. These include anatomical structures, defined by the spatial relationships between cells; cause-effect patterns, which capture how neurons *can* influence one another; and activation patterns, as measured through functional brain imaging techniques, to name just a few. But which kinds of neural structures should neurophenomenal structuralists focus on?

In this section, we delve deeper into the concept of exploitability by offering criteria to determine which neural structures are exploitable candidate NCCCs, and which should be dismissed.

### 4.1. Exploitable neural structure

To begin, we need to understand which relational neural configurations within the brain can potentially influence downstream processing in a structure-preserving—and, under a structuralist framework, content-preserving—manner. According to Shea (2018), an "exploitable neural structure" is one in which the relationships between parts of the neural structure can systematically influence and alter downstream processing. This understanding has important implications, as it allows us to introduce specific criteria that distinguish viable NCCC candidates from those that fail to meet even the most basic criteria.

### 4.1.1. Sensitivity Criterion

The Sensitivity Criterion simply asks whether changes in a neural structure influence downstream processing in any meaningful way. It is important to recognize that many neural properties—and the structures they form—do not contribute to the brain's information processing. For instance, the color of neurons, though detectable, does not play a role in neural communication because the brain lacks mechanisms to utilize these properties as signals for communication between different regions, so it would not matter for downstream processing if we painted cells with rainbow colors.

Hence, the first question that a neurophenomenal structuralist should ask is: "can a given kind of neural structure impact downstream processing?". As a case study, let us consider anatomical organization—the spatial layout of neurons in a population. The brain has intricate anatomical organization, both between and within brain regions. For example, early visual cortex displays orientation selectivity maps, where cells detecting visual edges of similar orientations are placed close together. This organization can be modeled as resulting from an interplay between statistics of the world and learning objectives (Lu et al. 2023), or as resulting from wiring length minimization (Chklovskii and Koulakov 2004).

However, in principle, this spatial arrangement could be changed without any impact on downstream processing. As Shea (2018, p.120) puts it: "connectivity not location of the cell is what counts". In line with this, the two prevailing perspectives on neural coding—those focusing on neural circuits (Sherringtonian) and neural state spaces (Hopfieldian)—do not consider spatial distances in their analysis. Instead, they prioritize the connectivity and functional profiles of neurons over their physical positioning (Barack and Krakauer 2021).

Another example is retinotopy—the fact that neurons are spatially arranged in the visual cortex according to the location of their receptive field in the world. This is due to well understood connectivity patterns in the retina, lateral geniculate nucleus, and visual cortex. Clearly, if one was to spatially move a neuron while keeping the connectivity pattern intact, then the neuron would not "notice", and its responses and all downstream processing would be unaffected. Hence, even though there is correspondence between the spatial location of a neuron in the brain and the spatial location of its receptive field, the spatial position in the brain is not causally relevant for downstream processing.

The Sensitivity Criterion states that, for a neural structure to be a good candidate NCCC, there must not just be correspondence with quality spaces, but a direct causal influence. The key implication here is that if phenomenal structure were directly mirrored by the anatomical arrangement of neural maps, such spatial rearrangements would result in unreportable changes in subjective experience. This would undermine the function of report-based quality spaces as reliable proxies. Therefore, anatomical structure cannot serve as the foundational basis for mirroring phenomenal structure if it can be dissociated from reported experiences so readily.

In summary, the Sensitivity Criterion asserts that for a neural structure to qualify as a viable NCCC candidate, downstream processes must be responsive to changes within that structure. This criterion is crucial in our effort to pinpoint viable NCCCs, as it filters out candidate structures whose relational configurations lack any causal relevance to the generation of subjective reports.

### 4.1.2. Organization Criterion

Building upon the Sensitivity Criterion—which establishes the requirement for a neural structure to exert some causal influence on downstream processes—the Organization Criterion takes this a step further by demanding that this influence be systematic and structured. As Shea (2018, p.121) notes, an "exploitable structural correspondence also requires that relation[s] should make a *systematic* difference to downstream processing". This means that for a neural structure to qualify as a valid NCCC, its influence must not only exist but also align with the principles of structural representation: similar vehicles yield similar contents.

To elucidate this concept, we can draw upon Godfrey-Smith's (2017) distinction between organized and non-organized representations. Unlike symbolic or nominal representations, which lack any systematic alignment between syntactic (vehicle) and semantic (content-related) properties, structural representations are characterized by their organized nature, where these properties are at least partially aligned. Consider, for instance, Paul Revere's lantern signals used to alert the colonial militia of the British approach: "one if by land, two if by sea." In this non-organized system, the number of lanterns holds no inherent semantic significance—the difference between "one" and "two" is purely arbitrary. However, if we were to use the brightness of a lantern to signal the size of the approaching force, this would be an organized system. In this organized representation, similar vehicles (degrees of brightness) map onto similar contents (sizes of the army) in a semantically significant manner.

Shea (2023) highlights the benefits of organized systems, explaining that they are more efficient, better suited for novel situations, and more resilient to errors. In an organized system, slight deviations in the signal don't cause significant changes in interpretation. For example, a small change in the bee dance does not drastically alter the interpreted distance to the nectar, ensuring the system is robust against noise. In contrast, non-organized systems, like language, are highly error-sensitive; changing "Tomato" to "Tolato" makes the word meaningless, while changing it to "Tornado" drastically changes its meaning.

The Organization Criterion challenges theories such as Integrated Information Theory (IIT), which posit that conscious experience is mirrored in the cause-effect structure of the brain, including active as well as inactive components and other dispositional properties (Tononi and Koch 2015). This broad approach does not align with the Organization Criterion because it implies that changes to the brain's causal structure can affect experiences without affecting downstream processes at all. In effect, changes in the causal structure do not systematically correspond to changes in the reported structure.

This can happen in several ways (Fig. 5). First, adding local neural loops that don't affect downstream processing can change the causal structure of a system without altering downward processes. Second, consider inhibiting silent cells—those neurons that are inactive in the current state. This alters the brain's causal structure without influencing current processing as well (Bartlett 2022), such that these inhibitions never translate into a difference in reported similarities either. Third, removing silent connections—those that do not currently participate in neural activity—also changes the cause-effect structure of the brain without impacting downstream processing. These examples demonstrate that not all parts of the cause-effect structure contribute to the organized representations underlying quality spaces (see Doerig et al. 2019; Herzog, Schurger, and Doerig 2022; Tsuchiya et al. 2019 for further discussion on the dissociation between causal structure and behavior).

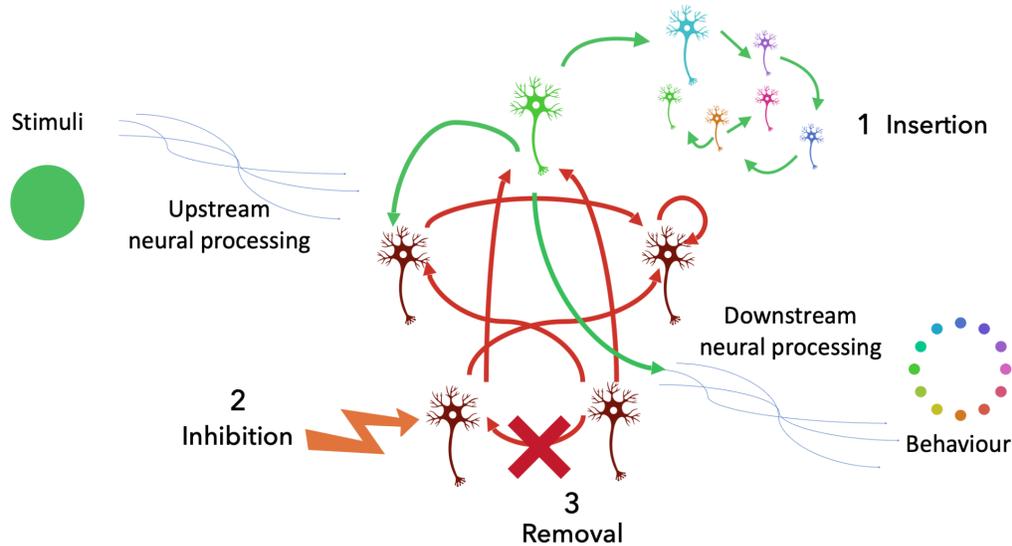

**Figure 5**: Illustration of dissociations between cause-effect space and quality space. 1.) Insertion of Downward Irrelevant Loops: Adding neural loops that don't affect downstream processing but change the causal structure. 2.) Inhibition of Silent Cells: Inhibiting inactive neurons changes the causal structure without affecting ongoing processes. 3.) Removal of Silent Connections: Removing inactive connections changes the causal structure but not immediate behavior. Red arrows depict inactive connections and green arrows active connections.

In conclusion, the Organization Criterion requires that for a neural structure to be considered a valid NCCC candidate, it must be organized such that changes in the structure predictably and systematically influence reported experiences. Small changes in the structure should result in small differences in content, while larger changes should lead to larger differences. If alterations in a neural structure do not influence downstream processing or do so but not in an organized manner, that structure should be excluded as an NCCC.

### 4.1.3. Interim Conclusion

In our search for viable NCCCs, we have excluded anatomical structures because they do not causally impact downstream processing, and causal structures because they do not impact downstream processing in an organized manner. Therefore, these kinds of structures are not properly exploitable and can easily be dissociated from the structure of the quality spaces as derived from behavior.

**4.2. Exploiting Activation Spaces**

While some structuralists focus on causal structure (Oizumi et al. 2014; Tononi and Koch 2015; Tononi et al. 2016), they are a minority. A widely used option for neurophenomenal structuralists is to suggest that the structure of *neural activation patterns* mirror phenomenal structures. Essentially, if two experiences are phenomenally similar, their corresponding activation patterns in a region of interest (ROI) should also be similar, while dissimilar experiences should produce more distinct patterns (Kriegeskorte and Kievit 2013; Cichy et al. 2019).

Brouwer and Heeger (2009) provide a key example of this activity-based neurophenomenal mapping approach. They used principal component analysis (PCA) to identify structural similarities between activation spaces of different ROIs and the psychophysical color space, finding that the activation patterns in area V4 most closely matched the circular structure of the color quality space, more so than in other areas like V1. Fink et al. (2021) suggest that such degrees of correspondence could help pinpoint potential NCCCs. The idea is to apply this approach to many ROIs to figure out which brain region (here, V4) best fits a behaviorally-measured quality space (here, color space). The winner is probably the NCCC for this phenomenal category (here, color).

However, this does not ensure that the brain actively exploits the relations between these activation patterns. This limitation is well recognized in NCC research; Aru et al. (2012) point out that the pre- and post-processes surrounding the true NCCs can inherit their correlational profiles. This challenge extends to structural correspondences as well. Indeed, the study by Brouwer and Heeger (2009) shows structural correspondences across multiple ROIs, not just one, which motivates the use of the degree of correspondence as a soft criterion. Importantly, there is no reason to assume that the ROI with the highest correspondence must be the only one that is actually exploited.

Hence, we agree with Fink et al. (2021) that the degree of correspondence can be a helpful soft criterion for identifying potential NCCCs. However, we believe that more decisive criteria are needed in addition—specifically, criteria that focus on whether the activation space of a neural region is actively exploi*ted* (as opposed to just being exploita*ble*) and plays a functional role in shaping reported experiences.

**4.2.1. Exploitation Criterion**

The key issue motivating the next criterion is well captured by Shea (2014): "In many cases where there is a salient isomorphism, the correspondence is not in fact exploited." The Exploitation Criterion addresses this by asserting that for a neural structure to qualify as an NCCC, its relational configuration must be utilized *structurally* by the brain's downstream processes. This criterion goes beyond mere structural correspondence; it requires that the brain genuinely engages with the relational information encoded within the activation space of an ROI.

Shea repeatedly (2014, 2018, 2023) emphasizes that not all organized representations qualify as structural representations because they are not exploited appropriately by downstream processes. For example, in the honeybee nectar dance, there is a clear correspondence between the number of wiggles and the distance to nectar sources—fewer wiggles indicate closer nectar. However, this system is not a structural representation because bees do not compare multiple dances to infer distances between different nectar sources; they respond to each dance independently. Thus, while there is organization within the nectar dances—similar vehicles yield similar content—the supervening relational structure is not exploited in a way that would make it a structural representation.

Shea (2023) also illustrates this with the example of neuronal spikes, explaining that "the spikes emitted by two neurons may be more or less similar in terms of the exact timing at which spikes are emitted (synchrony). That kind of similarity will be causally irrelevant unless there are downstream processes that can detect and respond to synchrony (which there may well be)." In other words, even if neurons fire in synchrony, if the brain's downstream systems do not utilize that synchrony, it has no relevance for the overall processing.

When we apply this concept to our case, we see that even if activation spaces in an ROI correspond to a given quality space, this correspondence alone does not guarantee that the ROI functions as an effective NCCC. Without appropriate exploitation, similarities between neuronal firing patterns will not contribute to the production of behavior and corresponding quality spaces.

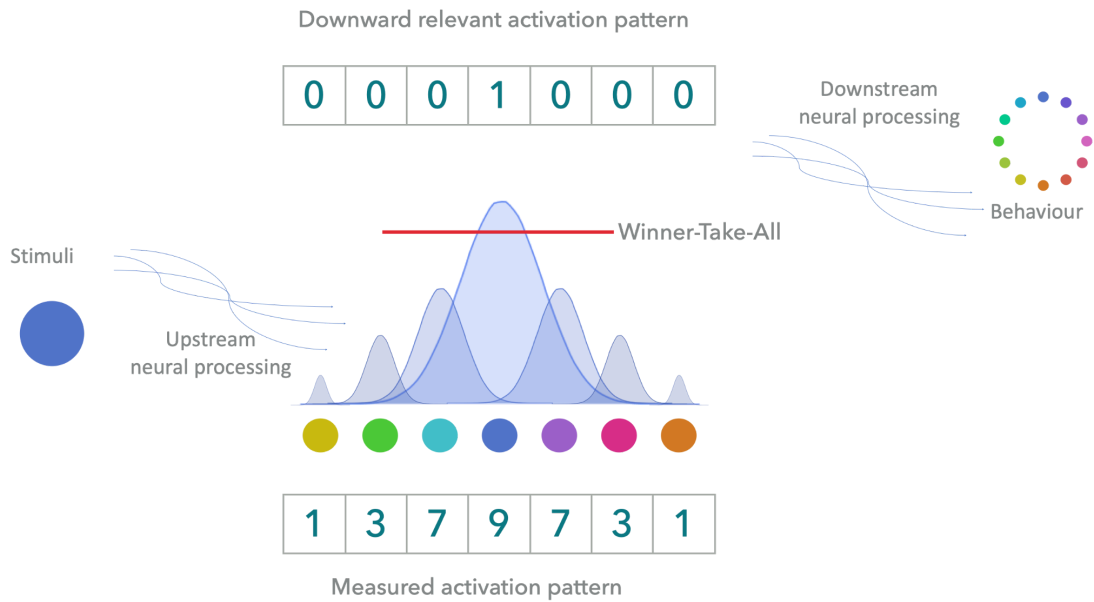

**Figure 6**: Illustration of 'Winner-Take-All' read-out
The figure shows a color-coding neural map reacting to a blue stimulus. Downstream neural processing employs a 'winner-takes-all' strategy, filtering out all but the strongest signal and reducing the rich, overlapping activation pattern to a simple binary signal. In this process the nuanced relational information about color similarities in the measured activation pattern is lost.

Take, for instance, a 'winner-takes-all' read-out mechanism that processes input from a neural map corresponding to color space (Fig. 6). In such a map, activation patterns for similar colors overlap systematically—similar colors produce more similar activation patterns than dissimilar ones. However, if the downstream processing systems reduce these complex patterns to a single dominant output, silencing or ignoring all other neural signals, this 'winner-takes-all' strategy would completely eliminate the rich relational information about color similarities present in the measured activity patterns.

As a result, instead of preserving the nuanced relationships between colors, each color would be represented by the distinct firing of one neuron, effectively stripping away the structured relationships inherent in the initial activation pattern. For instance, now the signal for orange would be just as different from the signal for red as it is from blue, rendering all color signals equally dissimilar and equidistant. Such a 'winner-takes-all' strategy has actually been proposed in color processing recently (Zaidi and Conway 2019) and thus presents a real caveat for areas like V4—though corresponding to color space—to act as a structural representation and thus as an NCCC.

This highlights that even though researchers can decode structural information from neural activity patterns, this does not necessarily imply that the brain itself uses this information in a structure-preserving or functionally meaningful way (Poldrack 2006; Klein 2010; Brette 2018; Ritchie et al. 2019). For a neural structure to be a valid NCCC, it's not enough to just contain decodable structural information—this information must be actively used by the brain's downward processes to shape subjective reports.

In summary, the Exploitation Criterion requires that downstream processes actively use the relational information within activation patterns. If not, the neural structure cannot be a viable NCCC candidate, as its relational configurations cannot be translated into the required structure-preserving behaviors underlying quality space theory.

### 4.2.2. Contextualization Criterion

The Contextualization Criterion emphasizes the need to consider the broader computational context when determining how neural structures relate to content. As Shea (2018, p.96) points out: "A vehicle that carries correlational information about one state of affairs will usually carry information about many. Different downstream systems may be interested in different pieces of information: different correlations may be of use to each" (Shea 2018, p.96) This criterion is crucial because it clarifies that equivalent activation spaces can also convey different meanings depending on how they are embedded in the overall processing of the brain.

Shea (2018) illustrates the ambiguity of signals with several examples, showing that identical representational vehicles can convey different meanings depending on how downstream systems use them. For instance, a firefly's light signal indicates "mating" to conspecifics but signals "prey" to predators. Similarly, in the brain, the same motor signal that commands movement can also inform perceptual systems about impending motion. The motor system interprets this signal as a directive to move, while the perceptual system uses it to predict sensory feedback. This dual usage is essential for the smooth coordination and control of actions. These examples underscore that the content of neural signals is ambiguous and context-dependent, emphasizing the importance of understanding how downstream systems interpret these signals.

a. Color processing

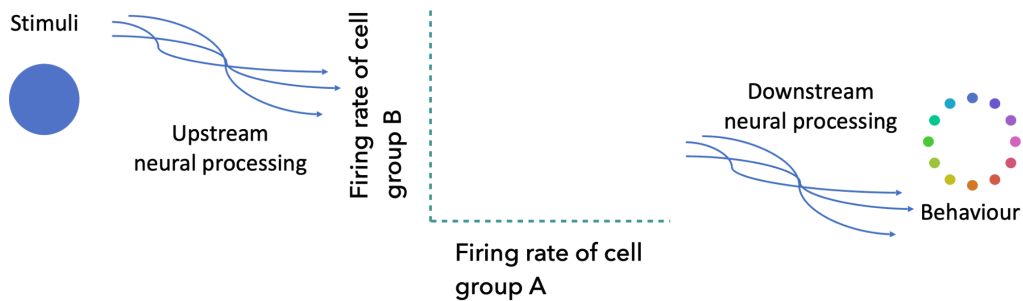

b. Emotional processing

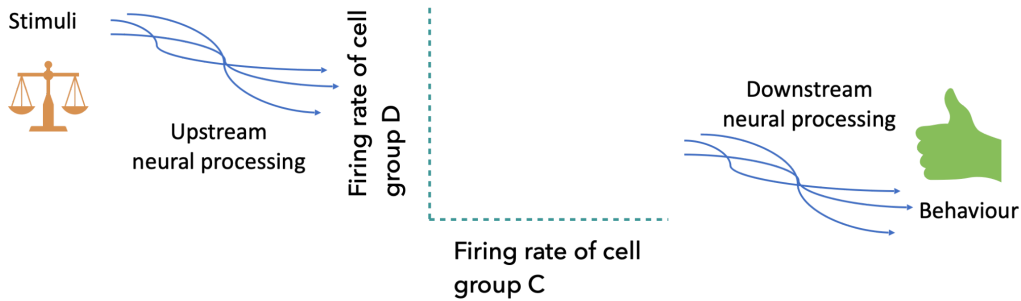

**Figure 7**: Illustration of the Ambiguity of Activation Spaces
The figure shows that two equivalent two-dimensional spaces can carry different contents depending on their context. In one case (a), the dimensions represent the opponent red-green and yellow-blue axes of the color space. In another case (b), the dimensions correspond to the opponent positive-negative valence and high-low arousal axes of affective processing. While both spaces share equal dimensionality, they are embedded within vastly different computational contexts—color perception versus emotional judgment.

This ambiguity extends to structural representations as well, since equivalent neural structures can be used for different things depending on their connections within the brain. For instance, consider two structurally equivalent two-dimensional activation spaces, one for color processing and another for affect, as Figure 7 illustrates. Both spaces are structured by two opponent processing streams, resulting in two bipolar dimensions. For color, these dimensions are Red-Green and Blue-Yellow (De Valois and De Valois 1993), while for affect, they are positive-negative valence and high-low arousal (Barrett and Russell 1999). Despite both being two-dimensional, their differing computational contexts—visual versus emotional processing—assigns these spaces entirely different representational roles.

This illustrates that the **nature of dimensions**—what each dimension in a neural space encodes—is determined not solely by the local structure of the activation space but critically by the computational context in which it is embedded. Without considering the broader computational processes, we cannot discern whether a particular dimension stands for a color axis like red-green or an emotional axis like positive-negative valence. Computational context must be added to locally encoded activation spaces to determine the nature of dimensions.

While the neural vehicles underlying these activation spaces may be drastically different, both embed a two-dimensional space, raising the possibility that the neural structures could, in principle, be exchanged. If properly connected to both upstream and downstream systems, these structures could switch roles, such that the neural vehicles previously involved in color perception take on the role of affect processing, and vice versa. This illustrates how the functional use of neural structures is not determined solely by their local vehicle properties but also by their broader computational integration, reinforcing the inherent ambiguity of local neural structures portrayed in isolation.

In summary, the Contextualization Criterion underscores the crucial role of computational context in determining the content of neural structures. Even if a local brain region's structure aligns with a quality space and is exploited, understanding its role within the brain's broader processing network is essential. Locally, without context, we cannot fully comprehend the representational role of an activation space, particularly its dimensions. This is why computational context is not merely an adjunct but a fundamental part of any structural NCCC.

## 5. Discussion

A key insight from our inquiry is that structural NCCCs must not only mirror phenomenal structures but also influence behavior in a way that preserves those structures. To address this, we introduced four criteria to filter out neural structures that only trivially correspond to quality spaces but are not exploited by downstream processes in an appropriate way.

The Exploitation and Contextualization Criteria, in particular, go beyond the current neurophenomenal mapping accounts by emphasizing the importance of the brain's broader computational network. For neural structures to qualify as viable NCCCs, they must be actively exploited by downstream systems, with this exploitation playing a significant role in content determination.

This analysis challenges Fleming and Shea (2024) by ruling out both local and rich global structuralist accounts. Local versions (e.g. Malach 2021) assume that local lateral connections in sensory areas 'implement' quality spaces for each sensory domain. This local conception is based on the idea that the activation patterns within these regions "exist in a similarity space, implemented in a neural activation space, the dimensions of which, as with size or colour, give the experience its particular character" (Fleming and Shea 2024). However, this conception raises a critical issue central to our Contextualization Criterion: how are these dimensions determined to provide that particular character?

We have shown that quality spaces cannot be implemented in such a fully local fashion. To determine the nature of an activation space's dimensions—that is, what each dimension in these spaces represents—it is essential to look at the broader computational context that exploits these locally encoded spaces. By ignoring the content-constituting role of computational context, local theories make the mistake of proposing NCCs that fail to even account for their content.

Similarly, Brette (2018) critiques the "neural coding" metaphor, which assumes that neural activity encodes stimuli locally. He argues that this approach relies on knowledge of the experimental conditions—information available to the experimenter, but not to the brain itself. The brain has to interpret its signals without access to their hidden causes. Equally, local structuralists—who infer the nature of an activation space's dimensions based on what experimental conditions triggered the activation patterns—make an inference that the neural system itself is not equipped to make. They thus prematurely assign content, overlooking the crucial role of computational context in determining the representational role of an activation space.

This critique also applies to rich global theories, as described by Fleming and Shea (2024), in which the global workspace simply replicates local activation patterns. This conception carries the same Content-Determination Fallacy by presuming that structural replication alone is sufficient to copy content. Moreover, because the nature of dimensions changes based on which local structures are replicated, rich global theories must account for how computational context adapts to accommodate for these shifts.

## 6. Conclusion

Neurophenomenal structuralism offers a promising approach to link neural structures and conscious experiences but faces significant empirical challenges. Our analysis shows that merely identifying correspondences between neural and phenomenal structures is insufficient. First, quality spaces—based on subjective reports—must be seen as behavioral proxies for phenomenal structures. Second, the mapping between these quality spaces and neural structures must go beyond trivial correspondence, focusing on exploitable neural configurations that actively shape the behaviors forming these quality spaces. Third, we introduced four exploitation-related criteria for identifying neural structures that play an active role in shaping these reports. Ultimately, these criteria demonstrate the critical role of computational context in determining the dimensions—and therefore the content—of activation spaces that purportedly implement quality spaces.

Acknowledging the importance of computational context challenges both local and rich global structural theories—as discussed by Fleming and Shea (2024)—which overlook how downstream processes determine the dimensions of activation spaces. This oversight leads to a content determination fallacy, meaning the proposed NCC fails to even fully account for the NCCC.

Our analysis thus offers clear criteria for rejecting theories and NCC candidates that fail to account for them. Future research should refine these criteria and explore whether these insights extend to other neural coding schemes or if valid local explanations for content-determination exist. This will further clarify the relationship between brain activity and consciousness, by enhancing our understanding of NCCCs.

As a final remark, this paper emphasizes the importance of addressing content-related issues when evaluating theories of consciousness. Neurophenomenal structuralism stands out by offering a clear hypothesis on how brain activity relates to experiences—proposing that similar brain patterns give rise to similar experiences. This precision is often lacking in other theories although crucial for their evaluation. We thus hope that all theories of consciousness one day provide an account of how their proposed NCCCs determine the structure of the experiences they are associated with.